\documentclass[12pt,preprint]{aastex}
 \makeatother

\def\vol#1  {{{#1}{\rm,}\ }}

\def\sec{\S }

\newcount\refno
\newcount\rfno
\def\eq{$^{\the\refno\ }$\advance\refno by 1}
\def\ad{\advance\rfno by 1}

\def\clock{\count0=\time \divide\count0 by 60
     \count1=\count0 \multiply\count1 by -60 \advance\count1 by \time
     \number\count0:\ifnum\count1<10{0\number\count1}\else\number\count1\fi}

\def\myputfigure#1#2#3#4#5%
{\hskip0.03\textwidth\vskip#5pt%\hfill
\makebox[0pt]{\hskip#2in
\includegraphics[width=#3\textwidth]{#1}}\vskip#4pt\hfill}

\def\Gcm2{\rm G~cm^2}

\def\beq{\begin{equation}}
\def\eeq{\end{equation}}
\def\bea{\begin{eqnarray}}
\def\eea{\end{eqnarray}}

\def \date         {\ifcase\month \message{zero} \or
                    January \or February \or March \or April \or May \or June
                    \or July \or
                    August \or September \or October \or November \or
                    December \fi
                    \space\number\day, \number\year}

\def\dv{{\delta v}}
\def\dl{{\delta \lambda}}
\def\kx{k^L_x}
\def\ky{k^L_y}
\def\kz{k^L_z}
\def\en{{\cal E}}
\def\rmin{{\cal R}_{c}}
\begin{document}

\title{Bypass to Turbulence in Hydrodynamic Accretion: Lagrangian Analysis of Energy Growth}
\author{Niayesh Afshordi\altaffilmark{1,2}, Banibrata Mukhopadhyay\altaffilmark{1,3}, and Ramesh Narayan\altaffilmark{1,4}}
\altaffiltext{1}{Institute for Theory and Computation, Harvard-Smithsonian Center for Astrophysics, MS-51, 60 Garden Street, Cambridge, MA 02138}
\altaffiltext{2}{nafshordi@cfa.harvard.edu}
\altaffiltext{3}{bmukhopadhyay@cfa.harvard.edu}
\altaffiltext{4}{narayan@cfa.harvard.edu}
% \accepted{ }

\begin{abstract}
Despite observational evidence for cold neutral astrophysical accretion disks, the viscous process which may drive the accretion
in such systems is not yet understood. While molecular viscosity is too small to explain the observed accretion efficiencies
 by more than ten orders of magnitude, the absence of any linear instability in Keplerian accretion flows is often used to
 rule out the possibility of turbulent viscosity.
 Recently, the fact that some fine tuned disturbances of any inviscid shear flow can reach arbitrarily large
  transient growth has been proposed as an alternative route to turbulence in these systems. We present an analytic study
  of this process for 3D plane wave disturbances of a general rotating shear flow in Lagrangian coordinates,
  and demonstrate that large transient growth is the generic
  feature of non-axisymmetric disturbances with near radial leading wave vectors. The maximum energy growth is slower than
  quadratic, but faster than linear in time. The fastest growth occurs for two dimensional perturbations, and is only limited by viscosity,
  and ultimately by the disk vertical thickness.

  After including viscosity and vertical structure, we find that, as a function of the Reynolds number, ${\cal R}$,
   the maximum energy growth is approximately $0.4 ({\cal R}/\log {\cal R})^{2/3}$,
  and put forth a heuristic argument for why ${\cal R} \gtrsim 10^4$ is required to sustain turbulence in Keplerian disks.
  Therefore, assuming that there exists a non-linear feedback process to replenish the seeds for transient growth,
  astrophysical accretion disks must be well within the turbulent
  regime. However, large 3D numerical simulations running for many orbital times, and/or with
  fine tuned initial conditions, are required to confirm Keplerian hydrodynamic turbulence on the computer.

\end{abstract}

\section{Introduction}\label{intro}

Accretion disks are very common in astrophysics.  Disks are found in
Active Galactic Nuclei (AGN), around newly forming stars
(proto-planetary disks), and surrounding compact stellar remnants
(white dwarfs, neutron stars, black holes) in binary systems
\citep[e.g.,][]{pringle}. Despite the overwhelming evidence for the
existence of accretion disks \citep[e.g.,][]{lin}, and decades of
literature dedicated to their understanding \citep[e.g., see][and
references therein]{bhr}, the detailed working mechanism of
accretion disks remains enigmatic at best.

One of the early puzzles in understanding the accretion phenomenon
was the clear inadequacy of molecular viscosity in driving accretion
in a Keplerian disk.  This led to the speculation of turbulence as a
proxy for viscous transfer of angular momentum \citep{weiz,shakura}.
The idea was particularly attractive because of the low viscosity of
astrophysical fluids (high Reynolds number $\sim 10^{10}-10^{14}$),
which, according to conventional wisdom, should make the shear flow
in accretion disks unstable to turbulence.

However, in the context of Keplerian disks relevant to most
astrophysical applications, no instability could be identified. The
linear instabilities that seemingly induce the onset of turbulence
in normal shear flows are stabilized by the Coriolis force
associated with rotation in a Keplerian disk.  Nonlinear
hydrodynamic simulations appeared to confirm the absence of
turbulence in such disks \citep[][ both hereafter referred to as
BHSW]{hbs,hbw}.

The breakthrough came in 1991, through re-discovery of the
Magnetorotational Instability \citep[MRI;][]{chandra} by Balbus and
Hawley \citep{bh}, who showed through Magnetohydrodynamic (MHD)
simulations that initial seed magnetic fields in an MHD disk grow
exponentially within a few rotation times, causing the onset of MHD
turbulence. The MRI instability is now widely accepted as the driver
of MHD turbulence in ionized disks \citep[see e.g.,][for a recent
review]{balbus}.

Despite the great success of the MRI in many astrophysical systems,
it is known that this instability will not operate in disks with
very small ionization fractions, where the magnetic flux is poorly
coupled to the gas.  Examples of systems with low ionization
fractions are proto-planetary disks, outer regions of AGN disks, and
white dwarf disks in the low state
\citep{gammie96,gammie98,fromang}.  The route to turbulence and
accretion in such neutral disks remains an outstanding puzzle in
theoretical astrophysics.

On the other hand, laboratory experiments of Taylor-Couette systems
(fluid flow between concentric rotating cylinders) seem to indicate
that, although Coriolis force delays the onset of turbulence, the
flow is ultimately unstable to turbulence for Reynolds numbers
larger than a few thousand \citep[e.g., see][]{richard2001}, even
for subcritical systems (systems with no linear instability).
\citet{long} reviews the experimental evidence for the existence of
turbulence in subcritical laboratory systems, and based on
phenomenological analogy, concludes that a similar process must
happen in astrophysical accretion flows. \citet{long} also claims
that the absence of turbulence in previous numerical simulations
(BHSW) is due to their small effective Reynolds number, which is
limited by the numerical viscosity caused by the finite resolution
of the simulation. Indeed, \citet{bech} see turbulence persisting in
numerical simulations of subcritical rotating flows for large enough
Reynolds numbers.

How does a shearing flow that is linearly stable to perturbations
switch to a turbulent state?  A possible explanation, known as {\it
bypass} transition, has been discussed in the fluid mechanics
community for some time \citep[see][and references therein, for an
overview]{grossmann,reshotko,schmid}, though its diffusion into the
astrophysical community has been slow \citep{ioan, chagelish,
tevzadze, yecko, umurhan}. The bypass concept is based on the fact
that {\it definite frequency} linear modes are not orthogonal in a
shear flow. Therefore, even if all of the linear modes are decaying,
a suitably tuned linear combination of them can still show an
arbitrarily large transient energy growth in the absence of
viscosity. In lieu of linear instabilities such as MRI, the
transient energy growth, supplemented by a non-linear feedback
process to repopulate the growing disturbances, could plausibly
sustain turbulence for large enough Reynolds numbers.
%{\bf
%(Niayesh, you tend to use a lot of hyphens in your writing.  I think
%the modern tendency is to eliminate hyphens as far as possible.
%Take a look at some recent ApJ articles to see what their current
%practice is.)}

This paper, along with a companion paper \citep[][ hereafter
MAN04]{bani}, investigates the transient growth of perturbations in
rotating shear flows, with an emphasis on applications to
astrophysical accretion disks. Both papers study the dependence of
the transient energy growth on various parameters of the system, in
particular the wave vector of the perturbations and the Reynolds
number. While MAN04 focuses on an eigenmode analysis in Eulerian
coordinates for a shearing flow restricted between rigid walls, the
present paper studies the bypass process in Lagrangian coordinates
for an infinite shearing flow. The advantage of the Lagrangian
approach is that there is no explicit coordinate dependence in the
equations, and therefore the solution can be decomposed into plane
waves.  The drawback, however, is the explicit time dependence of
the equations which prohibits definite frequency solutions, and thus
requires explicit integration in time for each mode. The two
analyses presented here and in MAN04 involve different
numerical/analytic techniques and are both, in our opinion,
valuable. The consistency of the results validates the general
picture of our understanding of transient growth in the linear
regime.

In \S2 we summarize the current understanding of the transient
growth phenomenon in shearing flows.  \sec\ref{eqs} introduces the
basic linear equations in Lagrangian coordinates in their most
general form. \sec\ref{ideal} constitutes the main body of the
paper, where we approach the problem of transient growth in inviscid
and incompressible flows for general plane wave solutions. This is
followed by \sec\ref{visc} and \sec\ref{cs}, where we study the
effects of viscosity and compressibility/vertical structure, respectively, and make the
connection to astrophysical accretion flows. Finally, in
\sec\ref{discuss} we discuss conditions for the emergence of turbulence, and its realization in numerical simulations.
 \sec\ref{conclude} summarizes our results and concludes the paper.

\section{Understanding Transient Growth through Swinging
Plane Waves}\label{transientsum}

As mentioned in \S1, the systematic approach to transient growth in
subcritical systems is through a linear combination of non-normal
decaying modes.  In particular, in a local region of an accretion
flow, different definite frequency modes with equal vertical and
azimuthal wave numbers, but different radial profiles, are
generically not orthogonal to one another, i.e. \beq \int d^3{\bf x}
~{\bf \dv}^{a}\cdot{\bf \dv}^{b} \neq 0,\eeq where ${\bf \dv}^{a}$
and ${\bf \dv}^{b}$ are velocity profiles of different modes.
Therefore, even though all the modes may decay with time, a solution
may still show a temporary energy growth for suitable initial
conditions because of the cross terms in the energy expression. For
a detailed description of the eigenmode approach, we refer the
reader to MAN04 and \citet{yecko}, which explain the numerical
methods used to find the maximum growth through optimizing a linear
combination of eigenmodes.

An alternative approach, which allows simple analytic treatment of
linear perturbations for the case of astrophysical accretion flows,
is the so-called {\it shearing box} approximation in which we study
the linear evolution of plane wave perturbations in Lagrangian
coordinates within a small region of an accretion flow
\citep{chagelish, tevzadze, umurhan}.  This is the approach we
follow in this paper.

Fig. \ref{vfield} shows our choice of local Cartesian coordinates:
$x$ is along the radial direction, $y$ is along the azimuthal or
streamwise direction, and $z$ is along the vertical direction.  The
unperturbed flow has a velocity in the $y$-direction and a velocity
gradient (shear) along the $x$ direction.  The Coriolis force
associated with rotation is described by an angular frequency vector
${\bf \Omega}$ pointed in the $z$ direction.  We define \beq q
\equiv -d\ln \Omega/d\ln R,\eeq which is a dimensionless parameter
that quantifies the shear in the local comoving box \footnote{
$q\Omega$ is twice the Oort constant $A$.}. For example, $q=3/2$
corresponds to a Keplerian accretion disk and $q=2$ to a disk with
constant specific angular momentum.

Starting with a plane wave in Lagrangian coordinates with radial
wave number $\kx$ (Eq. \ref{pwave}), we show in \sec\ref{eqs} that
the Eulerian radial wave number $k_x$ of the plane wave evolves as a
function of time $t$ according to: \beq k_x = \kx +(q\Omega t)k_y
,\eeq while the azimuthal and vertical wave numbers $k_y$ and $k_z$
remain unchanged.  Thus, the plane wave is effectively frozen into
the flow and is swung around by the shear.  For simplicity, as in
\citet{chagelish} and \citet{umurhan}, let us consider
`two-dimensional' ($k_z=0$) incompressible and inviscid
perturbations.
 %{\bf (Niayesh,
%Bani's paper refers to this as `two-dimensional' perturbations.
%Which is better, planar or two-dimensional?  We should use the same
%expression in both papers.)}
In this regime, in general, the
two-dimensional vorticity $\xi = k_x \dv_y - k_y \dv_x$ remains
constant, while the two-dimensional divergence vanishes due to
incompressibility, i.e. $k_x \dv_x + k_y \dv_y =0$.  Therefore, the
energy content of the plane wave scales as: \beq {\cal E} \propto
\dv^2_x + \dv^2_y = \frac{\xi^2}{k^2_x +k^2_y} =
\frac{\xi^2}{\left[\kx+(q\Omega t)k_y\right]^2+k^2_y},\eeq which
reaches a maximum when $k_x=0$, or $\kx/k_y=-q\Omega t$.  If the
initial wave vector $k_x=\kx$ is negative, then the maximum is
reached at positive time, and the energy growth factor is given by:
\beq G_{max} \equiv \frac{{\cal E}_{max}}{{\cal E}(0)} \simeq
(\kx/k_y)^2 = (q\Omega t)^2 ~{\rm for} ~-\kx/k_y \gg 1.
\label{gpg}\eeq As promised, the growth can become arbitrarily large
for long enough time. As we show in \sec \ref{cs}, the maximum
growth is limited only by viscosity and scales as ${\cal R}^{2/3}$,
where ${\cal R}$ is the Reynolds number \citep[][MAN04]{chagelish,
yecko}. Note that, after reaching the maximum, the linear solution
drops and decays to zero asymptotically. Therefore, to have
sustained turbulence, the perturbations must become non-linear
before reaching the peak and must provide sufficient feedback to
keep the perturbations going. \citet{umurhan} describe a 2D
numerical simulation of the local accretion flow in which they find
sustained turbulent behavior in the absence of viscosity.

Moving away from the $k_z=0$ plane, it is seen that for $ q \lesssim
2$ (the regime of interest for astrophysical disks), the maximum
growth drops \citep[][ MAN04]{yecko}. However, \citet{tevzadze}
argue that for $k_z \neq 0$, although the maximum growth is smaller,
vertical stratification may cause the solution to have a
non-vanishing asymptotic value, which is a fraction of the maximum
growth.  It is not clear if this will significantly help the onset
of turbulence.

%\citet{balbus04} presents a scaling symmetry, which maps small-scale %inviscid and incompressible solutions to a large-scale one. He then %uses this fact to argue that the absence of instabilities in %numerical simulations of \citet{hbs} and \cite{hbw}, justifies its %absence on any smaller scales, although they are not resolved by the %simulations, and therefore the onset of turbulence is not affected %by the numerical resolution. Although the first part of this %argument is certainly true, unlike exponential growth, the quadratic %transient growth

This concludes our summary of (the rather thin) astrophysical
literature on transient growth and the bypass mechanism. In the
following sections, we present a formal treatment of the Lagrangian
hydrodynamic equations in the shearing box approximation, and study
the transient growth of general plane wave solutions.

\section{Hydrodynamic Equations in the Local Lagrangian Coordinates (Shearing Box Approximation)}\label{eqs}

We limit the calculations to scales much smaller than the thickness
of the disk, which, for a geometrically thin disk, is in turn much
smaller than the distance to the central object.  We thus ignore the
boundaries.  Within a comoving local box, and in terms of Lagrangian
coordinates, the Navier-Stokes and continuity equations can be
written as: \bea \dot{\bf v} &=& -c^2_s {\bf \nabla} \lambda + \nu
\nabla^2 {\bf v} + 2 {\bf v \times \Omega}, \label{nv}\\
\dot{\lambda} &=& -{\bf \nabla.v},\label{cont}\\ \dot{\bf r} &=&
{\bf v}({\bf r}^L), \label{lag}\\ {\bf \nabla} &\equiv& {\bf
\partial r^L \over \partial r}{\bf . \nabla^L}, \label{grad} \eea
where ${\bf v}$ is the fluid velocity vector, ${\bf r}$ and ${\bf
r}^L$ are Eulerian and Lagrangian position vectors respectively,
$\nu$ is the kinematic coefficient of (molecular) viscosity, and
$\lambda \equiv \ln \rho$ is the logarithm of fluid density. Since
the local box rotates with the flow, we have a Coriolis term on the
right hand side of the Navier-Stokes equation (\ref{nv}), which is
proportional to the local angular velocity ${\bf \Omega}$; in the
shearing box approximation, the Coriolis parameter $\Omega$ is taken
to be independent of position.  We have dropped the centrifugal and
gravitational accelerations, as they cancel in the equilibrium flow
and do not contribute to the perturbation equations.

\begin{figure}
\includegraphics[width=0.75\linewidth, angle=-90]{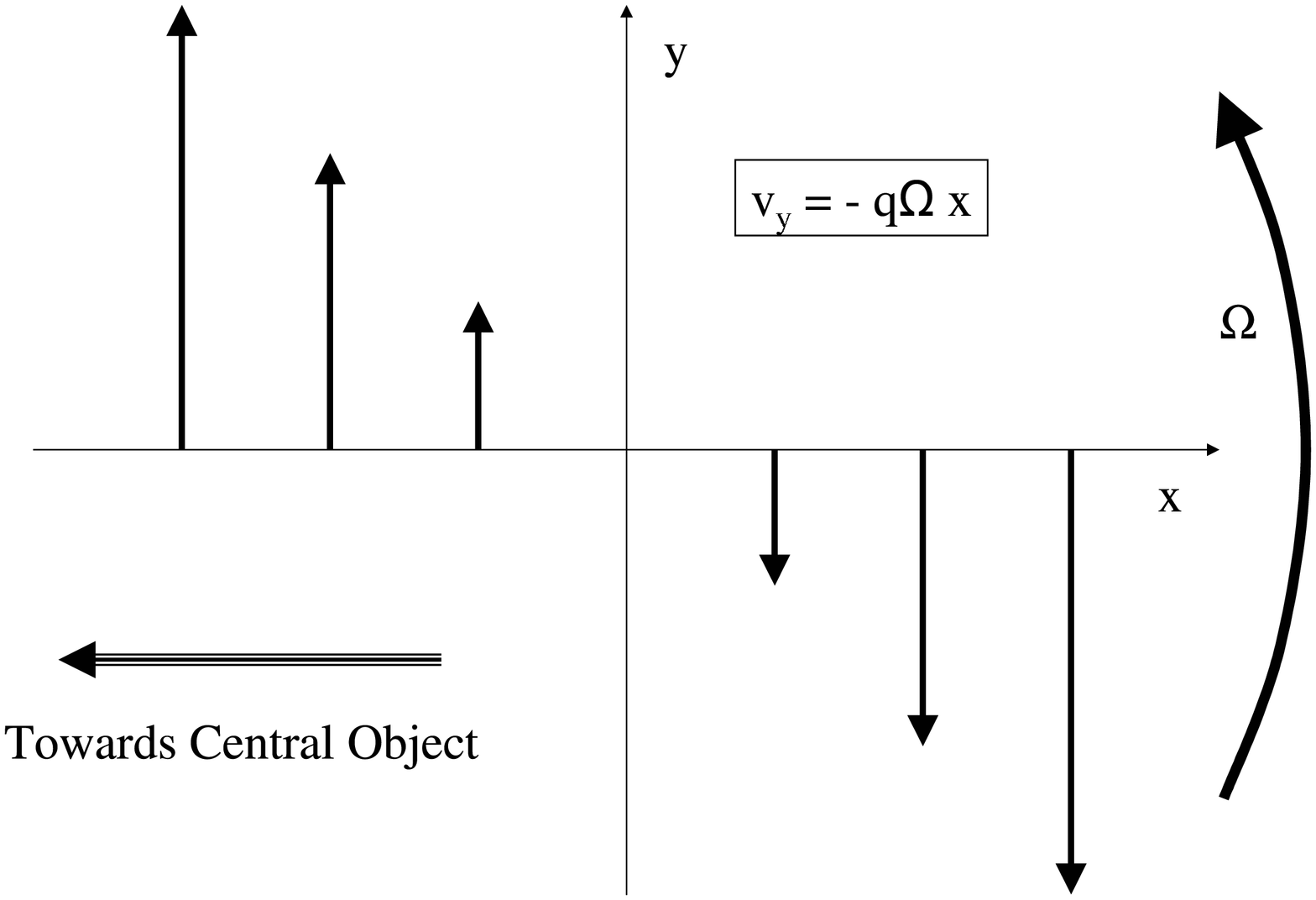} \caption{\label{vfield} A plot of the unperturbed flow in the local comoving box studied in this paper. The thick arrows represent the velocity field.} \end{figure}

Fig. \ref{vfield} shows the unperturbed velocity field within the
local comoving box.  Let us define $\Omega {\frak q}$ as minus the
gradient of the unperturbed velocity field, \beq \Omega{\frak q}
\equiv -{\bf \nabla v} = -{(\bf \nabla \Omega) \times R} = \left(
\begin{array}{ccc} 0 & q\Omega & 0 \\ 0 & 0 & 0
\\ 0 & 0 & 0 \end{array} \right);~~ q=-\frac{d\ln \Omega}{d\ln R} ,
\label{defq}\eeq where ${\bf R} = (R,0,0) $, and $R$ is the distance
to the center of the disk. Notice that, as the velocity field is
normal to its gradient, and flow lines are straight in the local
approximation, Eq. (\ref{lag}) can be integrated to give \beq {\bf
r}^L = {\bf r} + \Omega t{\bf r} .{\frak q} \Rightarrow {\bf
\partial r^L \over \partial r} = {\bf 1} + \Omega t {\frak q}, \eeq
and thus \beq {\bf \nabla} = ({\bf 1} + \Omega t {\frak q}){\bf .
\nabla^L} \label{nablal}. \eeq

In deriving the equations for {\it Eulerian} perturbations in
Lagrangian coordinates, we note that, as the unperturbed velocity
field ${\bf v}_0$ has a spatial gradient (Fig. \ref{vfield}), it has
a non-vanishing time derivative in the perturbed Lagrangian
coordinates, i.e.  \beq {\bf \dot{\dv}}= {\bf \dot{v}} - {\bf
\dot{v}}_0 = {\bf \dot{v}} - {\bf v.\nabla v}_0 = {\bf \dot{v}} +
\Omega {\bf v}.{\frak q}. \eeq Combining this with Eqs. (\ref{nv})
and (\ref{cont}) yields the linear perturbation equations: \bea
\dot{\bf \dv} &=& -c^2_s {\bf \nabla} \dl + \nu \nabla^2 {\bf \dv} +
2 {\bf \dv \times \Omega} + \Omega{\bf \dv}.{\frak q}, \label{dnv}\\
\dot{\dl} &=& -{\bf \nabla.\dv},\label{dcont}\eea where the Eulerian
gradients are related to the Lagrangian gradients according to Eq.
(\ref{nablal}). We note that since ${\frak q}$ is constant, the
Navier-Stokes and continuity equations have no explicit dependence
on the Lagrangian coordinates. Therefore, we can decompose a general
linear perturbation into plane wave solutions of the form: \beq {\bf
\dv}, ~ \dl \propto \exp(i{\bf k}^L.{\bf r}^L).\label{pwave} \eeq
Note that, although the Lagrangian fluid equations (\ref{nv} and
\ref{cont}) look linear in ${\bf v}$ and $\lambda$, the above plane
wave description breaks down for non-linear perturbations, since Eq.
(\ref{nablal}) will be modified at higher orders.

For a Lagrangian plane wave solution, the Eulerian wave number ${\bf
k}$ can be obtained from Eq. (\ref{nablal}): \beq {\bf k} =
(k_x,k_y,k_z)=({\bf 1} + \Omega t {\frak q}){\bf . k^L} = (k^L_x
+q\Omega tk^L_y, k^L_y,k^L_z) \label{kkl}.\eeq This can be combined
with Eqs. (\ref{dnv}) and (\ref{dcont}) to yield the mode equations
for Lagrangian plane wave solutions \bea \dot{\dv_x} &=& -i c^2_s
(\kx+\Omega tq\ky)\dl+2\Omega\dv_y-\nu k^2\dv_x, \label{dvx}\\
\dot{\dv_y}&=&-i c^2_s \ky\dl+(q-2)\Omega\dv_x-\nu k^2\dv_y,
\label{dvy}\\ \dot{\dv_z}&=&-i c^2_s \kz\dl-\nu k^2\dv_z,\label{dvz}
\\ \dot{\dl} &=& -i[(\kx+\Omega
tq\ky)\dv_x+\ky\dv_y+\kz\dv_z],\label{dl}\eea where $k$ is the total
Eulerian wave-vector \beq k^2 = (\kx+\Omega
tq\ky)^2+(\ky)^2+(\kz)^2. \eeq

For given values of $q,\Omega,\nu$, and $c_s$, Eqs.
(\ref{dvx}-\ref{dl}) provide a homogeneous set of linear first order
differential equations with time dependent coefficients. In the rest
of this paper, we attempt to study the growth of energy $\en$,
defined as \beq \en = \frac{1}{4}(|\dv^2_x|+|\dv^2_y|+|\dv^2_z|),
\eeq for different modes in various limits of the parameter space of
interest to astrophysical/physical problems. As we discussed in
\sec\ref{transientsum}, the presence of a large transient energy
growth, even in a small region of phase space, may act as a possible
trigger for the onset of self-sustained turbulence.

Eqs. (\ref{dvx}-\ref{dl}) can be simplified by introducing the 2D divergence and vorticity, $\Delta$ and $\xi$, defined as \bea \Delta &=& k_x \dv_x + k_y \dv_y, \\ \xi &=& k_x \dv_y -k_y \dv_x, \eea which yield \bea \dot{\xi}&=& (q-2)\Omega\Delta -\nu k^2 \xi, \label{dxi}\\ \dot{\Delta}&=&\left(\frac{2 q\Omega k_x k_y}{k_x^2+k^2_y}-\nu k^2\right)\Delta + 2\Omega\left(1 - \frac{qk^2_y}{k_x^2+k^2_y}\right)\xi-ic^2_s(k_x^2+k^2_y)\dl,\\
\dot{\dv_z} &=& -i c^2_s k_z\dl-\nu k^2\dv_z,\\ \dot{\dl} &=&
-i[\Delta+k_z\dv_z].\label{dkl}\eea In terms of the new variables,
$\en$ can be written as \beq {\cal E} =
\frac{|\Delta^2|+|\xi^2|}{4(k^2_x+k^2_y)}+\frac{|\dv^2_z|}{4}
\label{en0}.\eeq

\section{Idealized Inviscid and Incompressible Flow}\label{ideal}

The Reynlods number corresponding to molecular viscosity is
typically very large in astrophysical accretion disks. In this
limit, we can visualize a regime in which \beq (c_s t)^{-2} \ll k^2
\ll (\nu t)^{-1}, \eeq where $t$ is the characteristic time of the
transient growth. Within this regime, we may neglect viscous effects
($\nu \simeq 0$) and we may also consider the fluid to be
effectively incompressible, i.e. $\dl \rightarrow 0$ and $c^2_s
\rightarrow \infty$, while the pressure perturbation $c^2_s \dl$
remains finite.

We begin by rearranging Eqs. (\ref{dxi}-\ref{dkl}) to get: \bea
\dot{\xi}&=& (q-2)\Omega\Delta -\nu k^2 \xi,\label{xi2} \\
\dot{\Delta} &=& \frac{2q\Omega k_xk_yk_z^2}{k^2(k^2_x+k^2_y)}\Delta
+2\Omega\left(\frac{k^2_z}{k^2}\right)\left[1-{qk^2_y\over(k^2_x+k^2_y)}
+\right]\xi-\nu k^2 \Delta + i\left(k^2_x+k^2_y\over
k^2\right)(\ddot{\dl}+\nu k^2 \dot{\dl}),\label{delta2}\nonumber\\ \\
k^2_z(ic^2_s\dl)&=&\dot{\Delta}+\nu k^2 \Delta -i(\ddot{\dl}+\nu
k^2\dot{\dl}),\\ k_z\dv_z&=&-\Delta+i\dot{\dl}, \eea with the energy
given by \beq {\cal E} =
\frac{|\Delta^2|+|\xi^2|}{4(k^2_x+k^2_y)}+\frac{|\Delta|^2+|\dot{\dl}|^2}{4k^2_z}.
\label{en2}\eeq

In the incompressible and inviscid limit, we may
ignore the terms linear in $\dl$ and $\nu$ on the right hand sides
of Eqs. (\ref{xi2}-\ref{en2}).  Eqs.  (\ref{xi2}) and (\ref{delta2})
then take the form \bea \dot{\xi}&=& (q-2)\Omega\Delta, \label{xiii}
\\ \dot{\Delta} &=& \frac{2\Omega k_z^2}{k^2(k^2_x+k^2_y)}\left\{
qk_x k_y \Delta +[(1-q)k^2_y+k^2_x]\xi \right\}.\label{deltaii} \eea
As the coefficients in the above equations are real, we can without
loss of generality focus on real solutions, and thus the energy
becomes \beq {\cal E} =
\frac{k^2\Delta^2+k^2_z\xi^2}{4k^2_z(k^2_x+k^2_y)}, \label{enii}\eeq
where $k_x=k^L_x+(q\Omega t)k_y$, is the only time-dependent
component of the wave vector.

\subsection{Near two-dimensional Growth ($k_z\ll k_y, \kx$)}\label{pg}

Eqs. (\ref{xiii}-\ref{enii}) are significantly simplified in the
two-dimensional regime, $k_z = \dv_z =0$ (see Chagilishvili et al.
2003; Umurhan \& Regev 2004). In this limit, we find \bea \Delta &=&
\dl=0, \qquad \xi = {\rm const.},\\ {\cal E}_{2D} &=& {\xi^2\over
4(k^2_x+k^2_y)}={\xi^2\over 4[(\kx+q\Omega tk_y)^2+{k_y}^2]}. \eea
We have already seen in \sec\ref{transientsum} that this leads to a
maximum energy growth factor of $(q\Omega t)^2$ for $q\Omega
t=-\kx/k_y \gg 1$ (Eq. \ref{gpg}).

%which yields the growth function \beq G_{2D}(t)=\frac{{\cal %E}_{2D}(t=T)}{{\cal E}_{2D}(t=0)} = %\frac{(\kx)^2+{k_y}^2}{(\kx+qTk_y)^2+{k_y}^2}.\eeq

%For a fixed time $T$, which is much longer than the characteristic %shear time, i.e. $qT \gg 1$, one can easily see that the maximum %value of the growth function is \beq G_{2D,max}(T) \simeq (qT)^2 %\label{gpg}\eeq, which occurs for $\kx \simeq - (qT)k_y$.

%Therefore, we see that for an idealized incompressible and inviscid %shear flow, irrespective of the rotation $\Omega$, we can achieve an %arbitrarily large energy growth for planar perturbations. The %growth, however, is quadratic in time, which is significantly slower %than the usual exponential energy growth expected from positive %eigenvalue instabilities such as MRI.

Starting with $\Delta=0$ and $\xi=\xi_0={\rm const.}$ as the zeroth
order solution, to first order in $k_z^2$, Eq. (\ref{deltaii})
becomes \beq \dot{\Delta} \simeq \frac{2\Omega
k^2_z}{k^2_x+k^2_y}\left[1 - \frac{q
k^2_y}{k^2_x+k^2_y}\right]\xi_0,\eeq which can be integrated to give
\beq \Delta = \Delta_0 + \frac{k^2_z \xi_0}{k^2_y}\left[(2q^{-1} -1)
\tan^{-1} (k_x/k_y)-\frac{k_x k_y}{k^2_x+k^2_y}\right] + O(k^4_z).
\eeq Here $\Delta_0= O(k^2_z)$ and so it vanishes in the limit
$k_z\to 0$.  Plugging this back into Eq. (\ref{xiii}), yields $\xi$
to first order in $k^2_z$ \beq \xi
=\xi_0+(q-2)\Omega\Delta_0t-\frac{k^2_z\xi_0}{q^2
k^2_y}\left(2-q\right)\left[\left(2-q\right)\frac{k_x}{k_y}\tan^{-1}({k_x
\over k_y}) -\ln\left(1+{k^2_x\over k^2_y}\right)\right]. \eeq Now
we can go back to Eq. (\ref{enii}) to find the evolution of energy
with time. Focusing on the solution with $k_z=0$ and maximum growth
with $\kx = - (q\Omega t) k_y$, after some manipulations and
marginalizing over $\xi_0$ and $\Delta_0$, we end up with an
expression for the modified maximum growth \beq G_{max}(t)= (q\Omega
t)^2 \left[1-4 \left({k_z\over q k_y}\right)^2 \left(2 -q\right)
\ln(q\Omega t) +O\left(k^2_z/k^2_y\right)\right]. \label{gkz}\eeq We
see that, as expected (see \sec\ref{transientsum}), the growth
function decreases as we move away from the $k_y$ axis so long as $q
\lesssim 2$.

\subsection{Near Axi-Symmetric Growth ($k_y \ll k_z,\kx$)}\label{as}

Another special case which allows a simple analytic solution is the
case of axi-symmetric perturbations, where $k_y=0$. Since there is
no azimuthal dependence, the wave pattern is not swung round by the
unperturbed flow, and therefore, there is no explicit time
dependence in the equations (see Eq.  \ref{kkl}). As a result, Eqs.
(\ref{xi2}-\ref{en2}) simplify to \bea \dot{\xi} &=& -(2-q)\Omega
\Delta, \\ \dot{\Delta} &=& \frac{2\Omega k^2_z}{k^2_x+k^2_z}\xi,
\eea which allow a harmonic solution \bea \Delta &=& A\exp(-i\omega
t)+B\exp(i\omega t), \\ \xi &=&
i\frac{(q-2)\Omega}{\omega}[A\exp(-i\omega t)-B\exp(i\omega t)],
\eea where \beq \omega^2= \frac{2(2-q)\Omega^2
k^2_z}{k^2_x+k^2_z}.\eeq

We first note that for $q>2$ the frequency $\omega=\pm i|\omega|$ is
imaginary.  This means that the system has exponentially growing
perturbations, a reflection of the Rayleigh stability criterion
(according to which a flow with specific angular momentum decreasing
outward is unstable).  However, this regime is not of interest since
real disks cannot survive here.

Let us, therefore, consider $q < 2$, where stable circular orbits
are allowed and an accretion flow is possible.  Without loss of
generality, we can change the origin of time to eliminate the phase
difference between $A$ and $B$. The leftover phase will be
irrelevant for calculating the energy, and thus we can assume that
$A$ and $B$ are real. Plugging this into Eq.  (\ref{en2}), after
straightforward manipulations, we arrive at \beq {\cal E}_{AS} =
\frac{1}{8}(k^{-2}_x+k^{-2}_z)\left[(A^2+B^2)\left(4-q\right)+2q AB
\cos(2\omega t)\right],  \eeq where the subscript $AS$ identifies the axi-symmetric solutions.

Since ${\cal E}$ is a periodic function of time, the maximum growth
factor is equal to the ratio of its maximum to its minimum: \beq
G_{AS} ={(A^2+B^2)\left(4-q\right)+2AB\left|q\right|\over
(A^2+B^2)\left(4-q\right)-2AB\left| q\right|}, \eeq which is
maximized for $A=B$: \beq G_{AS,max} = {4-q+\left|q\right|\over
4-q-\left| q\right|} = \left(2\over 2 -q\right)^{{\rm
sgn}(q)}.\label{aximax}\eeq For most astrophysical disks, $q>0$,
i.e. the angular velocity decreases with increasing radius, and
therefore significant growth can only happen in the limit of a
constant angular momentum disk ($2-q = d\ln(\Omega R^2)/d\ln R \to
0$). For a Keplerian disk with $q=3/2$, the maximum possible energy
growth is only a factor of 4.

Focusing on the marginal case of a disk with constant angular
momentum ($q=2$), since the harmonic frequency $\omega$ is zero, we
find solutions that are linear in time: \beq \Delta = \frac{2\Omega
k^2_z \xi t}{k^2_x+k^2_z}, \qquad \xi = {\rm const.}. \eeq
The
growth is then given by \beq G_{AS}(t;q=2) =
1+\frac{k^2_z}{k^2_z+k^2_x}(2\Omega t)^2. \eeq As in the case of
two-dimensional perturbations, we see that the growth for
axisymmetric perturbations of a constant angular momentum disk is
quadratic in time and can be arbitrarily large in the absence of
viscosity.  In contrast, for any $q<2$, there is a firm limit to the
maximum growth, given by equation (\ref{aximax}).

Now, let us consider small but non-vanishing values of $k_y$. As a
small value of $k_y$ corresponds to a slowly changing frequency, we
can use the WKB approximation to solve Eqs.
(\ref{xiii}-\ref{deltaii}).  This gives \beq \xi(t) \simeq
\hat{\xi}(t) \exp\left[-i\int dt~ \omega(t)\right], ~~{\rm and}~~
\Delta(t) \simeq \hat{\Delta}(t) \exp\left[-i\int dt
~\omega(t)\right].\label{wkb} \eeq Ignoring terms of order $k^2_y$
in Eq. (\ref{deltaii}), we find after some simple manipulations that
\beq \dot{\Delta} = \Delta \frac{d}{dt}\ln(k^2_x/k^2)+\frac{2\Omega
k^2_z}{k^2} \xi \label{wkb_delta}.\eeq Plugging (\ref{wkb}) into
Eqs. (\ref{xiii}) and (\ref{wkb_delta}), and keeping only terms
linear in $k_y$, yields \beq i\omega(t) \hat{\xi} \simeq (2-
q)\Omega \hat{\Delta} ~~ {\rm ,and}~~ \hat{\xi}^2 \propto
\frac{k^2_x}{\omega k^2}. \eeq Plugging this result into the
expression for energy (Eq. \ref{enii}), and ignoring second order
terms in $k_y$, we find \beq {\cal E}_{\rm WKB} \propto k^{-1}
=(k^2_x+k^2_z)^{-1/2}, \eeq which reaches a maximum when $k_x = \kx
+k_y(q\Omega t) = 0$. Therefore, the maximum growth becomes \beq
G_{max}(t)= \sqrt{(\kx)^2+k^2_z\over k^2_z} \simeq (q\Omega
t)\left|\frac{k_y}{k_z}\right|\left[1+O\left(k^2_y\over
k^2_z\right)\right] ~~{\rm for}~~ q\Omega t \gg 1,\frac{k_z}{k_y}
\label{gwkb}.\eeq Thus, even near axi-symmetric plane-wave solutions
can achieve arbitrarily large transient growth.  However, the growth
is only linear in time, whereas it is quadratic for two-dimensional
solutions. Therefore, the growth is significantly slower than for
the two-dimensional case.

\subsection{Ideal Growth for a General Plane Wave Solution}\label{gpws}

Having analysed the regions near the two axes in wave vector space, we now
consider general perturbations of an incompressible and inviscid flow.
We have not been able to find an analytic solution to Eqs. (\ref{xiii}-\ref{deltaii}),
but we are able to put a lower bound on the growth.

%The idea is that the general solution a linear homogeneous equation
%like \beq {\bf \dot{X}} = {\bf A}(t) {\bf X}, \eeq where ${\bf X}$
%is a vector (e.g., $(\xi,\Delta)$) and ${\bf A}$ is a time-dependent %matrix, is given by \beq {\bf X}(t) = {\rm Texp}\left[\int_0^t dt
%{\bf A}(t)\right] {\bf X}(0). \label{solv}\eeq Although the
%time-ordered exponential in the above solution, in general, does not
%have a closed form, its determinant does, i.e. \beq {\rm det}~{\rm
%Texp}\left[\int_0^t {\bf A}(t) dt\right] = \exp\left[\int_0^t dt
%{\rm Tr} {\bf A}(t)\right]. \eeq

First, consider how the combination $\xi_1\Delta_2-\xi_2\Delta_1$
evolves with time, where $(\xi_1,\Delta_1)$ and $(\xi_2,\Delta_2)$
are two independent solutions of Eqs. (\ref{xiii}-\ref{deltaii}).
From these equations we obtain \beq
\frac{d}{dt}(\xi_1\Delta_2-\xi_2\Delta_1) = \frac{2q\Omega
k_xk_yk^2_z}{k^2(k^2_x+k^2_y)} (\xi_1\Delta_2-\xi_2\Delta_1),\eeq
which can be easily integrated to give \beq
(\xi_1\Delta_2-\xi_2\Delta_1)\left(k^2\over k_x^2+k^2_y\right) =
{\rm const.}\label{wrons}\eeq

Now consider two solutions with initial conditions
$(\xi_0,-\Delta_0)$ and $(\xi_0,\Delta_0)$, which have the same
initial energies: \beq {\cal E}_1(t=0) = {\cal E}_2(t=0)=
\frac{k_0^2\Delta_0^2+k^2_z\xi_0^2}{4k^2_z(k^2_0-k^2_z)}, \eeq where
\beq k_0^2 = (\kx)^2+k_y^2+k_z^2.\eeq After time $t$, the sum of the
energies of the two solutions is: \bea {\cal E}_1(t) +{\cal E}_2(t)
&=& \frac{k^2(\Delta_1^2+\Delta_2^2)+k^2_z
(\xi_1^2+\xi_2^2)}{4k^2_z(k^2_x+k^2_y)} =
\frac{(k^2\Delta_2^2+k^2_z\xi^2_1)+(k^2\Delta_1^2+k^2_z\xi^2_2)}{4k^2_z(k^2_x+k^2_y)}
\nonumber \\ &\geq& \frac{2 k k_z
(\xi_1\Delta_2-\xi_2\Delta_1)}{4k^2_z(k^2_x+k^2_y)} = (4\xi_0
\Delta_0) \left[4 k^2_z (k^2_0-k^2_z)\right]^{-1}(k^2_0 k_z /k),\eea
where we have used Eq. (\ref{wrons}) to substitute for
$\xi_1\Delta_2-\xi_2\Delta_1$ in terms of its initial condition.

We can now translate this result to a lower bound on the average
growth function of the two solutions: \beq \frac{G_1(t)+G_2(t)}{2}=
\frac{{\cal E}_1(t) +{\cal E}_2(t)}{{\cal E}_1(0) +{\cal E}_2(0)}
\geq \left(2 k_z k^2_0\xi_0\Delta_0/ k \over
k^2_0\Delta_0^2+k^2_z\xi^2_0\right). \eeq Maximizing this lower
bound over the initial conditions $(\xi_0,\Delta_0)$, and noting
that the average of $G_1$ and $G_2$ is a lower bound on the maximum
growth, we finally arrive at: \beq G_{max}(t) \geq {\rm Max}{k_0
\over k} \simeq (q\Omega t) \frac{k_y}{\sqrt{k^2_y+k^2_z}}, ~~{\rm
for}~ q\Omega t\gg 1,\frac{k_z}{k_y}. \label{lbound}\eeq The maximum
is achieved for $\kx \simeq - (q\Omega t)k_y$.

We first note that the above lower bound is smaller than the maximum
growth near the $k_{y}$ axis (Eq. \ref{gpg}), but becomes
asymptotically equal to the growth close to the $k_z$ axis (Eq.
\ref{gwkb}). More generally, this result implies that {\it
arbitrarily large transient growth is a generic feature of plane
waves with relatively large leading radial wave numbers} ($\kx \gg
k_y,k_z; \kx k_y <0$).

%We can now draw a picture of the growth function behavior in the %phase space of wave-vectors. We first note that the solutions to %Eqs. (\ref{xiii}-\ref{deltaii}) should only be a function of the %ratio of wave-vectors, e.g. $k_y/\kx$ and $k_z/\kx$. In \sec %\ref{as} we saw that, if $k_y=0$, for viable disk dynamical models %($q<2\Omega$), no significant growth can be found. However, %\sec\ref{pg} argued that, if $k_z =0$, solutions can have an %arbitrarily large growth. Here, we see that this large growth is %indeed a generic feature of the linear inviscid and incompressible %perturbations of an astrophysical disk, and the growth suppression %on the $k_z$ axis is merely an exception.

One interesting observation is how the slope of the energy growth
function depends on position in the $k_y-k_z$ plane.  We find: \beq
{d\ln G_{max} (t) \over d \ln(q\Omega t)} = \left\{
\begin{array}{ll} 1 & \mbox{if $k_y \ll k_z$}, \\ 2-
4q^{-1}\left(2q^{-1}-1\right) (k^2_z/k^2_y)& \mbox{if $k_z \ll
k_y$}, \end{array} \right.\label{sigmal}\eeq which is a result of
Eqs. (\ref{gwkb}) and (\ref{gkz}).  We see that, in general, the
growth function behaves as a power law in $q\Omega t$, i.e. $G
\propto (q\Omega t)^{\sigma}$, where $1\leq \sigma \leq 2$. We can
justify this conjecture by plugging a scaling ansatz into Eqs.
(\ref{xiii}-\ref{deltaii}): \beq \xi = \hat{\xi} k^{\beta}_x, \qquad
\Delta = \hat{\Delta} k^{\gamma}_x. \eeq This ansatz satisfies the
equations in the limit $k_x \gg k_y,k_z$, if \bea \beta &=&
\gamma+1, \\ \gamma(\gamma+1) &=&
-2q^{-1}(2q^{-1}-1)(k_z/k_y)^2,\eea or \beq 2\gamma = -1 \pm
\sqrt{1-8q^{-1}(2q^{-1}-1)(k_z/k_y)^2}. \eeq Plugging this result
into Eq. (\ref{enii}), we see that ${\cal E} \propto k^{{\rm
Re}(2\gamma)}_x$. Since the scaling solution breaks down when $k_x
\sim k_y,k_z$, the maximum growth relative to the initial value of
$\kx = -(q\Omega t) k_y$ takes the form \beq G_{max}(t) \simeq
\left[(q\Omega t) \left({k_y \over
\sqrt{k^2_z+k^2_y}}\right)F(k_z/k_y)\right]^{\sigma} ~{\rm for}~~~
(q\Omega t) \gg 1 \label{gmaxt}\eeq where \beq \sigma ={\rm
Max}~{\rm Re}(-2\gamma) = \left\{ \begin{array}{ll} 1 & \mbox{if
$k_z > \mu k_y $}, \\ 1+ \sqrt{1-(k_z/k_y)^2/\mu^2}& \mbox{if $k_z <
\mu k_y$}, \end{array} \right.\label{sigma}\eeq where \beq \mu = q
\left[8(2-q)\right]^{-1/2} \simeq 0.20 ~{\rm for}~ q=3/2. \eeq

We note that the above result for the logarithmic slope of the
maximum growth function is consistent with our previous asymptotic
results for the two limits $k_y \ll k_z$ and $k_z \ll k_y$ (Eq.
\ref{sigmal}). The exact value of the dimensionless factor
$F(k_z/k_y)\sim 1$ does not come out of the scaling argument.
However, based on our solution in the large and small $k_y/k_z$
limits (Eqs. \ref{gwkb} and \ref{gkz}), we can say that $F$ goes to
$1$ as its argument goes to zero or infinity. Moreover, the lower
bound in Eq. (\ref{lbound}) requires that $F(x)\geq 1$ for $x\geq
\mu$.

\begin{figure}
\includegraphics[width=\linewidth]{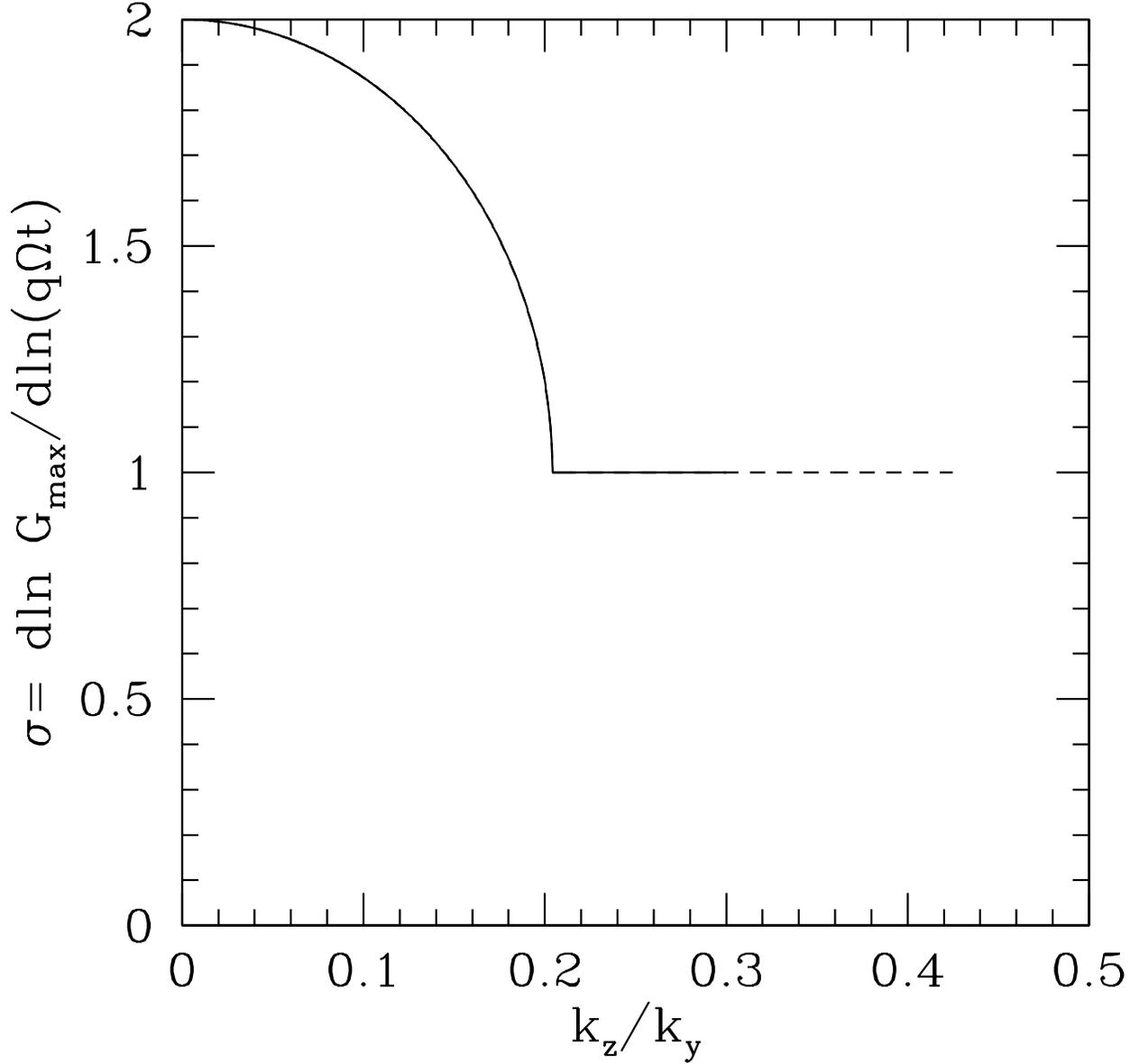}
\caption{\label{sigma_fig} The logarithmic slope of the maximum
growth function, $\sigma$, i.e. $G_{max} \propto (q\Omega
t)^{\sigma}$, plotted as a function of $k_z/k_y$ for a Keplerian
disk. For larger values of $k_z/k_y$, $\sigma$ remains constant.}
\end{figure}

Fig. \ref{sigma_fig} shows how the logarithmic slope of the growth
function, $\sigma$, depends on $k_z/k_y$ for a Keplerian accretion
flow. For $k_z<\mu k_y$, the slope monotonically increases with
decreasing $k_z$, reaching its maximum when $k_z=0$, as pointed out
by \citet{tevzadze,yecko}; and also MAN04.  Thus, the fastest growth is achieved for
two-dimensional solutions.  For $k_z > \mu k_y$, as for the $k_z \gg
k_y$ case (\sec\ref{as}), the solution is oscillatory with a growing
amplitude, yielding a maximum energy growth which is linear in time.

\section{Dependence of the Maximum Growth on Viscosity}\label{visc}

In this section, we study the effect of viscosity on the growth of
incompressible local perturbations of an accretion disk. Neglecting
the terms of order $\dl$ in Eqs. (\ref{xi2}-\ref{delta2}), but
keeping the terms that include viscosity, we end up with \bea
\dot{\xi}+\nu k^2 \xi &=& (q-2)\Omega\Delta, \label{xiiv} \\
\dot{\Delta}+\nu k^2 \Delta &=& \frac{2\Omega
k_z^2}{k^2(k^2_x+k^2_y)}\left\{ qk_x k_y \Delta +[(1-q)k^2_y+
k^2_x]\xi \right\}.\label{deltaiv} \eea Now, introducing \beq
\tilde{\xi} = \xi \cdot \exp\left[\nu \int k^2 dt\right], {~\rm
and}~~ \tilde{\Delta} = \Delta \cdot \exp\left[\nu \int k^2
dt\right], \eeq we see immediately that $\tilde{\xi}$ and
$\tilde{\Delta}$ satisfy the same equations that $\xi$ and $\Delta$
satisfy (Eqs. \ref{xiii}-\ref{deltaii}) for inviscid perturbations.
As the expression for the energy (\ref{enii}) remains unchanged, we
see that in the presence of viscosity, the growth function is simply
modified to \beq G_{max}(t) = G_{max}(t;\nu=0) \exp\left[-2\nu
\int^t_0 k^2 dt^{\prime}\right]. \label{gvisc}\eeq

For $q\Omega T \gg 1$, based on the results of the last section, we
know that the inviscid maximum growth is proportional to $(q\Omega
t)^{\sigma}$, where $1\leq \sigma \leq 2$, and the maximum is
achieved when $k_x=\kx +(q\Omega t)k_y \simeq 0$. Plugging these
into the above expression for the modified growth yields \beq
G_{max}(t) \propto (q\Omega t)^{\sigma} \exp\left[-\frac{2\nu
k^2_y}{3q}(q\Omega t)^3\right],\eeq which reaches a maximum for:
\beq q\Omega t_{max} = \left(\sigma q \Omega \over 2\nu
k^2_y\right)^{1/3}. \label{tmax}\eeq Plugging this back into Eq.
(\ref{gvisc}), we find that \beq G_{max}= G_{max}(t_{max};\nu=0)
\cdot \exp(-\sigma/3), \eeq where $G_{max}(t_{max};\nu=0)$ was obtained
in Eq.  (\ref{gmaxt}).

Consider as an example the case of two-dimensional perturbations,
where the solution has a closed form and the inviscid growth is
equal to $(q\Omega t)^2$.  In the presence of viscosity we find \beq
G_{max}(k_z=0) \simeq \left(q \Omega \over \nu k^2_y\right)^{2/3}
\exp\left(-2/3\right). \label{gmaxnp}\eeq

\section{Vertical Structure and Finite Speed of Sound}\label{cs}

The heuristic argument for assuming incompressibility in the
analysis so far is that, if the wavelength of the velocity
perturbations is much shorter than the sound horizon for the time of
interest, then the density perturbations (i.e. sound waves) reach
equilibrium early on and thus the density is effectively uniform
during the timescale of interest for velocity perturbations.  For a
geometrically thin disk around a gravitating mass, the vertical
half thickness of the disk $H$ is comparable to the sound horizon
corresponding to one disk rotation time \citep[e.g.,][]{pringle}:
\beq H \sim c_s \Omega^{-1}.\eeq Therefore, for processes that take
{\it longer} than one rotation time, wavelengths {\it shorter} than
the disk thickness can be approximately treated as incompressible.

We can refine this heuristic picture by focusing on the
two-dimensional perturbations ($k_z=0$) discussed in \sec \ref{pg},
since their solutions are available in closed form, and solving Eqs.
(\ref{dxi}-\ref{dkl}) to first order in $c^{-2}_s$. As we have
already discussed the effect of viscosity in the last section, we
will for simplicity ignore viscosity in the following analysis.
Assuming $k_z=\dv_z= \nu=0$, Eqs. (\ref{dxi}-\ref{dkl}) can be
simply combined to give \bea \Delta &=& -{\dot{\xi} \over
(2-q)\Omega}, \\ \dl &=& {i(\xi-\xi_0) \over (2-q)\Omega}, \\
\ddot{\xi} &=& \left(\frac{2q\Omega k_xk_y}{k^2_x+k^2_y}-\nu
k^2\right)\dot{\xi}+2\Omega^2(q-2)\left(1-{qk^2_y \over k^2_x
+k^2_y}\right)\xi +c^2_s(k^2_x+k^2_y)(\xi_0 -\xi), \eea where
$\xi_0$ is a constant.

Recognizing that $\xi=\xi_0$ and $\Delta=\dl =0$ are the solutions
to zeroth order in $c_s^{-2}$, we can easily write down the first
order solutions: \bea \xi &\simeq& \left[1+
\frac{2\Omega^2(2-q)}{c^2_s(k^2_x+k^2_y)}\left(1 - {qk^2_y \over
k^2_x+k^2_y}\right)\right] \xi_0, \\ \dl &=&
\frac{2i\Omega\xi_0}{c^2_s(k^2_x+k^2_y)}\left(1 - {qk^2_y \over
k^2_x+k^2_y}\right), \\ \Delta &=&
\frac{4q\Omega^2k_xk_y\xi_0}{c^2_s
(k^2_x+k^2_y)^3}\left[(2q-1)k^2_y- k^2_x\right].\eea Noting that the
maximum growth occurs when $k_x =0$, we find the correction to the
growth function: \bea G_{max}(t) &=& (q\Omega t)^2 \left\{
1-2(2-q)(q-1)(c_s k_y/\Omega)^{-2}+ O(c_s k_y/\Omega)^{-4}\right\}
\nonumber \\ &\simeq& (q\Omega t)^2 \exp\left\{-2(2-q)(q-1)(k_y
H)^{-2} +O(k_y H)^{-4}\right\},\label{cor_cs}\eea where $H \equiv
c_s/\Omega$.

%Although we need to know the correction for $k_y H \sim 1$ to find
%the effect on the maximum growth, let us use Eq. (\ref{cor_cs}) to
%estimate the maximum growth in the presence of both viscosity and
%compressibility/vertical structure. This formula has also the right
%asymptotic behavior for $k_y H \ll 1$, as different parts of the
%disk oscillate independently with the epicyclic frequency, and thus
%the growth remains small.  {\bf (Niayesh, I don't understand this
%paragraph)}

At this point, we should note that the vertical structure of the
disk puts a lower limit on the vertical component of the wave vector
$k_{z,min} \sim \pi/(2H)$ (assuming free boundary conditions at
$z=\pm H$)\footnote{Since the $z$-dependent parameter that appears
in the equations is $c_s$, $H$ is the characteristic scale for
variations in $c_s$, and not the un-perturbed density. In
particular, for an isothermal disk ($c_s=$ const.), $k_z$ has no
minimum, even for a disk with finite thickness. However, for a
realistic geometrically thin and optically thick disk, $c_s$ is
expected to drop significantly at the disk surface, which puts a
lower limit on the vertical wave number $k_z$.}. Therefore the
pre-factor of $(q\Omega t)^2$ must be replaced by $(q\Omega
t)^{\sigma}$, where $\sigma$ is given in Eq. (\ref{sigma}), and is
smaller than $2$ for a finite value of $k_z$. Let us also define the
Reynolds number ${\cal R}$ by \beq {\cal R} \equiv {\Omega H^2 \over
\nu} . \eeq Modifying Eq.  (\ref{cor_cs}) for $k_z=\pi/(2H)$, and
using Eq. (\ref{tmax}) we arrive at: \bea \ln G_{max}(k_z={\pi \over
2H}) &\simeq& \nonumber\\ {2\over 3} [1-2(2-q)q^{-2} {\left({\pi/
2}\right)^2} &{(k_y H)^{-2}}& ]\ln\left[q{\cal R} (k_y
H)^{-2}\right] -{2\over 3}-2(2-q)(q-1)(k_y H)^{-2} +
O(k_yH)^{-4}.\nonumber\\\label{lngmax}\eea

Due to the non-algebraic dependence of $G_{max}$ on $k_y$, one
cannot express the value of $k_y$ that maximizes $G_{max}$, or
$G_{max}$ itself, in a closed form. However, we can find an
asymptotic expansion, in the limit of $\ln {\cal R} \gg 1$. In this
limit, we find \bea G_{max} &\simeq& \left[2\exp(-1)q^3\over
(2-q)\pi^2\right]^{2/3} \left({\cal R}\over\ln {\cal
R}\right)^{2/3}\cdot\exp\left[O\left(\ln\ln{\cal R}\over \ln{\cal
R}\right)\right] \nonumber \\ &\simeq& 0.36 \left({\cal R}\over\log
{\cal R}\right)^{2/3}\cdot\exp\left[O\left(\ln\ln{\cal R}\over
\ln{\cal R}\right)\right] ,~ {\rm for ~a ~Keplerian ~disk~}( q=3/2).
\eea This maximum is achieved for \bea (k_y H)^2 &\simeq&
\frac{\pi^2(2-q)}{2q^2}\ln{\cal R} + O(1) \nonumber\\ &\simeq& 2.5
\log{\cal R} +O(1),~ {\rm for~a ~Keplerian ~Disk~}( q=3/2).\eea

For example, to reach a maximum growth of $\sim 10^3$, we need a
Reynolds number $\sim 10^6$, and the maximum growing perturbations
have $k_yH \simeq 4$, $k_zH\sim \pi/2$. We note that, in this
regime, ignoring the terms of order $(k_yH)^{-4}$, as we did in Eqs.
(\ref{cor_cs}-\ref{lngmax}), introduces $\lesssim 1\%$ error, and is
therefore justified.

\section{Discussion}\label{discuss}

Throughout this paper, we have demonstrated that arbitrarily large
transient growth of plane wave perturbations in a stable cold disk
is possible in the restricted phase space of modes with relatively
large radial leading wave numbers (i.e. $k_x \gg k_y,k_z; k_x k_y
<0$). This growth is only limited by viscosity, and eventually by
the vertical thickness of the accretion disk, as we saw in
\sec\ref{visc} and \sec\ref{cs}.

So why do 3D simulations of local hydrodynamic accretion flows fail
to see the onset of turbulence for a Keplerian disk (BHSW)? As we
mentioned in \sec\ref{intro}, one possibility is the limited
numerical resolution of the 3D hydrodynamic simulations
\citep[e.g.,][]{long}. \citet{balbus04} invokes the scale invariance
symmetry of inviscid Navier-Stokes equations, to argue that the lack
of any instability on large scales probed by simulations implies
stability for smaller scales. Although this argument may hold for
positive eigenvalue instabilities such as MRI, the amplitude of
transient growth is directly affected by the dynamical range of the
simulation (i.e. the effective Reynolds number). Therefore, the
resolution of a simulation may be a critical factor which decides if
an energy growth large enough to sustain turbulence can, or cannot
be achieved.

A yet more important factor in simulating the bypass phenomenon may
be the assumed initial conditions of the simulations. Although the
steady turbulent phase is expected to be independent of initial
conditions, due to the restricted nature of modes with large
transient growth, the time needed to reach the steady turbulent
phase will significantly depend on the choice of initial conditions.
For example, based on the results of \sec\ref{gpws}, we can show
that, starting with an isotropic distribution of energy in (3D)
k-space, the total linear energy decays in the linear regime.

In order to see this, let us consider the general case of a
$D$-dimensional isotropic initial energy distribution within the
phase space of modes, i.e. $d{\cal E} = f(k) {\bf d^Dk}$. Here,
$D=3$ corresponds to a 3D isotropic initial condition, $D=2$ is a 2D
isotropic distribution with $k_z=0$, and $D=1$ refers to initial
conditions with $k_y,k_z \ll k_x$. Since, the Lagrangian modes are
orthogonal, the maximum energy growth is the sum of the maximum
energy growth of the individual modes, i.e. \beq \bar{G}_{max} =
{\int G_{max}({\bf k}) f(k) {\bf d^Dk} \over \int f(k) {\bf d^Dk}}.
\eeq At time $t \gg \Omega^{-1}$, the maximum growth is peaked at
$-k_y = (q\Omega t)^{-1} \kx$, and $k_z=0$ with $G_{max} \sim
(q\Omega t)^2$. Based on the analyses of \sec\ref{gpws}, we can also
roughly estimate the width of the peak, i.e. $\Delta k_y \sim
(q\Omega t)^{-1} k_y \sim (q\Omega t)^{-2} k_x^L$ and $\Delta k_z
\sim k_y \sim (q\Omega t)^{-1} \kx$. In particular, for isotropic 3D
initial conditions, we find \beq \bar{G}_{max,3D} \sim {\int
(q\Omega t)^{2} f(\kx) (q\Omega t)^{-3} (\kx)^2 d\kx \over \int
f(\kx) (\kx)^2 d\kx} = (q\Omega t)^{-1}, \eeq implying that, despite
the presence of a large transient growth $\propto t^2$, as a result
of the shrinking phase-space volume (which decays as $\Delta k_y
\Delta k_z \propto t^{-3}$) the total energy in the perturbations
decays as $t^{-1}$, which is completely consistent with the BHSW
results.

In 2D, as it was recently pointed out by \citet{johnson1}, the
shrinking of the phase space volume exactly cancels the transient
growth, which implies a constant energy and thus, lack of any
significant growth in the linear regime: \beq \bar{G}_{max,2D} \sim
{\int (q\Omega t)^{2} f(\kx) (q\Omega t)^{-2} \kx d\kx \over \int
f(\kx) \kx d\kx} = 1, \eeq which is also consistent with simulations
of \citet{umurhan}.

Therefore, it is only with near one-dimensional initial conditions,
i.e. $\kx \gg k_y,k_z$, where we can obtain significant transient
growth of $\bar{G}_{max,1D} = (q\Omega t)^2$ until $\kx = -(q\Omega
t) k_y$.

One may speculate that in a 3D simulation the total energy decays
until/unless it is re-distributed into near radial leading modes
through non-linear couplings. BHSW follow the evolution for only a
few orbital times, which is probably not enough to reach the
expected turbulent phase with near-radial structure. However, a
direct way to test the viability of this scenario is to study a 3D
simulation with near-radial initial conditions, and see if a
self-sustained turbulent phase can be realized.

 A more subtle question to address is the minimum energy growth (or Reynolds
number) required for the onset of sustained turbulence. Unlike
linear evolution, the answer to this question requires solving
non-linear equations, which can be usually done only through
numerical methods. For example, \citet{umurhan} see sustained
turbulence throughout the duration of their inviscid 2D simulation,
while their run with a ${\cal R} \sim 10^5$ decays in a few hundred
rotation times. However, it can be formally shown that due to the
steady decay of comoving vorticity in 2D, one cannot sustain
turbulence in a periodic box without external forcing. Therefore 2D
simulations will not be able to pinpoint $\rmin$, the critical
Reynolds number necessary for the onset of sustained turbulence.
However, this does not necessarily pose a problem for real disks, as
the modes that show the largest growth are inherently
three-dimensional (see the end of \sec\ref{cs}).

Assuming that there exists a non-linear feedback process to
repopulate the growing disturbances, let us present a heuristic way
for estimating $\rmin$. Most theoretical studies of 2D turbulence
are based on describing the turbulent flow as a gas of 2D
interacting vortices \citep[see e.g.,][and refrences
therein]{miller}. While in isotropic turbulence the vortices are
typically circular, as can be seen in simulations such as that of
\citet{umurhan}, in a shear flow, the vortices are stretched along
the stream (our $y$ direction). The aspect ratio of vortices in
their simulation of a 2D Keplerian flow is within the range $6-11$.
Let us now {\it conjecture} that this number is intrinsic to
Keplerian 2D turbulence, and identify it with the minimum value for
the ratio $\kx/k_y$. Note that the latter is $q\Omega t_{max}$ at
the time of maximum growth, which is limited by viscosity through
Eq. (\ref{tmax}). The rationale of our argument is that if the
energy growth for modes with $\kx/k_y = 6-11$ is significantly
hindered by viscosity, then vortices of the corresponding scale damp
and thus turbulence does not develop at smaller scales. Now,
defining the effective Reynolds number for a simulation box of size
$L$ as ${\cal R} = \Omega L^2/\nu,$ and noting that the minimum
value of $k_y$ available to the modes is $2\pi/L$ for periodic
boundary conditions, Eq. (\ref{tmax}) for a 2D Keplerian flow gives
\beq \rmin \sim \frac{(2\pi)^2}{1.5}\left(\kx \over k_y\right)^3
\sim 6\times 10^3 - 4\times 10^4. \eeq

Based on analogy with the onset of turbulence in subcritical Couette
flow, MAN04 find a value for $\rmin$ that is consistent with the
high end of the above range. We thus conclude that the critical
Reynolds number required for sustaining turbulence in 3D numerical
simulations of a local Keplerian flow is (at least) $\sim 10^4$,
which is comparable with the effective dynamical range in BHSW
simulations. Therefore, we expect larger 3D simulations, with
appropriately chosen initial conditions, and/or long enough run time
to start seeing the possible emergence of Keplerian turbulence.

Here, we note that as the turbulent vortices are stretched in the
streamwise direction, more resolution in the radial (compared to the
streamwise) direction is needed to resolve the turbulent structure.
Therefore, the most efficient way to simulate the turbulent regime
is to look at a simulation box that is elongated in the streamwise
(y) direction by a factor of $6-11$ compared to the radial (x)
direction, while similar number of (asymmetric) cells in x and y
directions are used. This will insure that the largest (i.e. least
damped) sheared vortices can fit in the simulation box, while the
vortices are adequately resolved in both directions. We do not have
a solid answer for the optimum vertical (z) size of the simulation
box but note that, due to the decay of vorticity in 2D, the vertical
structure is necessary to sustain the turbulence.

\section{Conclusions}\label{conclude}

In this paper, for the first time, we present a three-dimensional
analytic study of the transient growth of energy for plane wave
disturbances of a rotating shear flow, in the limit of large speed
of sound and small viscosity, and consider its direct application to
astrophysical accretion disks. We see that, although the growth is
fastest for 2D disturbances, a large transient energy growth is a
generic feature of non-axisymmetric disturbances with relatively
large radial leading wave numbers ($k_x \gg k_y,k_z; k_x k_y <0$).
The maximum growth is quadratic in time for 2D disturbances, but
becomes linear in time for relatively large vertical wave numbers
($k_z \gtrsim k_y$), and long times.

After including the effects of viscosity and compressibility/finite
disk thickness, we show that the maximum energy growth scales as
$({\cal R}/\log {\cal R})^{2/3}$, where, ${\cal R} \gg 1$, is the
Reynolds number. Therefore, for neutral astrophysical accretion
disks with ${\cal R} \sim 10^{10}-10^{14}$, assuming that there
exists a non-linear feedback process to repopulate the growing
disturbances, the transient growth can act as an alternative to the
more conventional MRI instability in ionized disks, in starting and
sustaining the turbulence necessary to explain the observed
accretion efficiencies.

Finally, we present a heuristic argument based on the aspect ratio
of sheared turbulent vortices to show that the critical Reynolds
number for sustaining Keplerian hydrodynamic turbulence in a
periodic shearing box should be $\sim 10^4$. A hydrodynamical
simulation needs to start with a fine tuned initial condition or
last many orbital times to reach the sustained turbulent regime.

\acknowledgments We would like to thank the anonymous referee for
helpful comments and suggestions.

 \end{document}